\documentclass[prb,showpacs,superscriptaddress,twocolumn,notitlepage]{revtex4-1}

\usepackage{amsmath}
\usepackage{amsfonts}
\usepackage{graphicx}
\usepackage{multirow}
\usepackage{chngcntr}
\usepackage{bm}

\usepackage{ulem}
\usepackage[usenames,dvipsnames]{xcolor}

\usepackage{natbib}
\usepackage[final=true]{hyperref}
\hypersetup{
   colorlinks=true,
    linkcolor=Maroon,
   filecolor=magenta,      
    urlcolor=cyan,
    citecolor=PineGreen,    
}

\newcommand{\m}{\mathbf}

\newcommand{\be}{\begin{eqnarray}}
\newcommand{\ee}{\end{eqnarray}}
\newcommand{\nn}{\nonumber}

\begin{document}


\title{Raman peak shift and broadening in crystalline nanoparticles with lattice impurities}


\author{S.~V.~Koniakhin$^*$}
\affiliation{Center for Theoretical Physics of Complex Systems, Institute for Basic Science (IBS), Daejeon 34126, Republic of Korea}
\affiliation{Basic Science Program, Korea University of Science and Technology (UST), Daejeon 34113, Korea}
\email{kon@ibs.re.kr}

\author{O.~I.~Utesov}
\affiliation{Center for Theoretical Physics of Complex Systems, Institute for Basic Science (IBS), Daejeon 34126, Republic of Korea}

\author{A.~G.~Yashenkin}
\affiliation{The Faculty of Physics of St. Petersburg State University, Ulyanovskaya 1, St. Petersburg 198504, Russia}
\affiliation{Petersburg Nuclear Physics Institute NRC ``Kurchatov Institute'', Gatchina 188300, Russia}

\begin{abstract}
  
The effect of point-like lattice impurities on nanoparticle Raman spectra (RS) is studied using both numerical and analytical methods. Particular cases of replacement atoms of various masses, vacancies, and disorder in interatomic bonds are considered.  It is shown that the disorder leads not only to the broadening of optical phonon lines but also to the shift of the corresponding Raman peak. The latter can be either positive (i.e., blueshift) or negative (redshift) depending on the type of impurities. Thus there is an additional contribution to the well-known redshift that occurs due to the size-quantization (confinement) effect. Considering nanometer-sized diamond particles as a representative example, we show that the broadening and the shift are, as a rule, of the same order of magnitude. The results are discussed in the framework of the self-consistent T-matrix approach. It is argued that both effects should be considered for accurate treatment of experimental Raman spectra. Simple recipes to do so are formulated for several important cases including NV centers in nanodiamonds.

  
\end{abstract}


\maketitle

\section{Introduction}

Nanoparticle research has been one of the most actively developing domains in material science in the last three decades. The achieved progress has its origins not only in discoveries of new synthesis methods but in the development of nanoparticle characterization techniques as well. Along with chemical composition and phase determination, the physical parameters of nanoparticles must be measured and standardized for commercialization and usage in industry.  This requirement is actual for carbon nanoparticles including nanodiamonds manufactured with the use of various synthesis methods. One of the most important nanoparticle parameters is their size. It can determine morphology~\cite{shery2018size,shames2022size}, mechanical~\cite{popov2017raman}, thermal~\cite{neeleshwar2005size,kidalov2010grain,ekimov2022size}, optical ~\cite{neeleshwar2005size,chung2007particle,koniakhin2020evidence,makino2022effect} and electronic~\cite{norris1996measurement,williams2010size} properties important for nanotechnology and material science. Moreover, bio-compatibility~\cite{shmakov2018investigation} and penetration efficiency~\cite{hsiao2014size,zhu2013size,kong2015size,kurdyukov2019fabrication} are also size-dependent which is important for chemistry and biology. The methods for size measurements are presented by local microscopy techniques (scanning electron~\cite{sedov2011gas}, transmission electron~\cite{neeleshwar2005size,williams2010size,dideikin2017rehybridization,shery2018size,makino2021straightforward}, and atomic force~\cite{muravijova2018atomic} microscopies) and global methods that measure the properties of nanoparticles arrays. The latter include dynamic light scattering~\cite{johnson1981laser,aleksenskii2012applicability,koniakhin2015molecular,koniakhin2018ultracentrifugation,koniakhin2020evidence} as well as small-angle X-Ray and neutron scattering~\cite{ozerin2008x,avdeev2009aggregate,tomchuk2015structural}.Broadening of crystalline peaks in powder X-Ray diffraction patterns~\cite{makino2021straightforward,koniakhin2018ultracentrifugation} also allows determination of the typical crystal size due to the finite size effects leading to the momentum conservation relaxation.

Raman peak of crystalline nanoparticles is redshifted and asymmetrically broadened with respect to one of the bulk materials~\cite{richter1981one,campbell1986effects,meilakhs2017new}. It happens because phonon levels in nanoparticles are quantized in the same manner as, for example, electron and exciton ones are in quantum dots~\cite{yoffe1993low}. The effective wave vectors of such optical phonons are defined by the standing wave geometry for a nanoparticle as a resonator. Their estimation yields $q \approx 2 \pi/L$ (for a particle of size $L$) and thus the optical phonon energy in nanoparticles is lower than the energy of the phonons in a bulk which usually corresponds to the Brillouin zone center. The general structure of the Raman spectrum (RS) of a single particle (which can hardly be measured) is a superposition of multiple narrow peaks from different optical vibrational modes resembling the Raman spectra of fullerenes, see, e.g., Fig. 1 from Ref.~\cite{ourDMM}. However, in practice, Raman spectra are measured for the ensemble of nanoparticles, for instance, in the dried powder. The scatter in nanoparticle size in an ensemble smears the comb-like structure of single particles Raman spectrum and provides the visible broadening or at least one of the contributions to the broadening.

The widely used Phonon confinement model (PCM)~\cite{richter1981one,campbell1986effects,osswald2009phonon,korepanov2017carbon} explained the Raman peak downshift and broadening as an effect of momentum conservation relaxation: various momenta contribute to the Raman spectrum of nanoparticles powder. Mathematically it was formulated as a smooth in real-space Gaussian envelope for optical phonons amplitude in a single nanoparticle with the boundary amplitude controlled by a certain parameter. Due to the Fourier transform properties it leads to similar exponential broadening in the momentum space and as a result in phonon energies. This reasoning is not fully physical because it leads to the appearance of a continuous distribution of phonon momenta in a single nanoparticle while in reality the vibrational states in nanoparticle are quantized like electron levels in quantum dots or vibrations in fullerenes~\cite{snoke1993bond}, diamondoids~\cite{jenkins1980raman,filik2006ramanDiamondoids} or small crystalline particles~\cite{ourDMM,filik2006ramanAbinitio,cheng2002calculations,zhang2005signature}. Another physically grounded meaning can be found for this envelope in reciprocal space and this meaning is the result of statistical averaging over the nanoparticle ensemble with different sizes and, possibly, lattice disorder realizations. Thus the previously discussed~\cite{zhu2005size,zhu2005size,grujic2009use} parameter $\alpha$ of the model absorbs all mechanisms of broadening including size and shape variation~\cite{ourEKFG,ourBench}, crystalline lattice impurities~\cite{our3,our4,ourSurf}, anharmonic effects~\cite{faraci2009quantum,chaigneau2012laser,dohcevic2009anharmonicity} and inhomogeneous laser heating~\cite{dorfler2022generalized}.

The previously developed combined dynamical matrix method and bond polarization model (DMM-BPM)~\cite{ourDMM} approach calculates the nanoparticle Raman spectra on microscopical grounds: dynamical matrix method for lattice vibrations and bond polarization model for calculating Raman tensor for each vibrational mode obtained with DMM. Repeating the calculations for various sizes allows for obtaining the Raman spectrum for a real powder with a given size distribution. The EKFG approach~\cite{ourEKFG} is its continuous formulation, in other words, the elasticity theory for optical phonons applied for nanoparticles Raman spectra. All the approaches above (DMM-BPM, EKFG, and PCM) require the phonon line broadening (instrumental or caused by the finite phonon lifetime) as a parameter. Previously, scattering on lattice impurities has been proposed as a main mechanism of the broadening~\cite{our3,our4}, and the corresponding theory has been successfully employed for the experimental Raman spectra description~\cite{ourBench}.

In the present study, we show that the nanoparticle lattice disorder in the form of point-like impurities not only provides the broadening of the phonon lines and, correspondingly, of the Raman spectrum but also results in the Raman peak shift accompanying the one stemming from the size-quantization effect. Importantly, this shift can be either red or blue depending on the type of disorder. In more detail, we perform a comprehensive study of various types of lattice impurities, including weak Gaussian disorder,  heavy and light replacement impurities, vacancies, and typical for diamond NV centers. We observe that in all cases the phonon line broadening is accompanied by the shift of the Raman spectrum. We use suitable theoretical methods (e.g., self-consistent T-matrix approach~\cite{utesov2021self}) for the discussion of the results and provide simple formulae for the ratio of the shift and the broadening in various observed regimes which turn out to be of the same order of magnitude for nanometer-sized particles.

The rest of the paper is organized as follows. In Sec.~\ref{STheor} we provide important details of the theoretical framework including analytical results for Bose particles scattering off impurities and a description of our numerics. Section ~\ref{SRes} is devoted to the comprehensive study of various types of the disorder. The obtained results are summarized and discussed in Sec.~\ref{SDisc}. Finally, we present our conclusions in Sec.~\ref{SConc}.

\section{Theory}
\label{STheor}

\subsection{Preliminary remarks}
\label{SModel}

To derive vibrational modes of molecules, atomic clusters, and nanocrystallites the following generic Hamiltonian in harmonic approximation can be used
\begin{equation}\label{eq_hgen0}
  {\cal H} = \sum_l \frac{p^2_l}{2 m_l} +  \frac{1}{2} \, \sum_{l l^{\prime}} K_{ll^{\prime}} \left( \mathbf{r}_l - \mathbf{r}_{l^{\prime}} \right)^2 + \mathrm{A.B.},
\end{equation}
where the designation A.B. stands for valence angle bending terms. Index $l$ spans all of $N$ atoms in the nanoparticle/nanocrystallite, $m_l$ is the mass of $l$-th atom and $K_{ll'}$ is its bond stretching rigidity while $l'$ being its neighbor. From this Hamiltonian, the equations of motion for all atoms can be directly obtained and after the Fourier transform in the time domain one can write the following dynamical matrix method (DMM) equation with  $3N\times3N$ matrix $D$:
\begin{equation}
  \label{eq_DynMat}
  m\omega^2 r_{p}=D_{pp^\prime}r_{p^\prime}\equiv\sum_{p^\prime=1}^{3N} \frac{\partial^2 {\cal H}(...~,~r_p~,~...,~r_{p^\prime}~,~...)}{\partial   r_{p} \partial r_{p^\prime}}r_{p^\prime},
\end{equation}
where index $p=3\cdot(l-1)+j$ denotes the displacement of $l$-th atom along three possible directions in Cartesian coordinates ($j=1,2,3$ are for x,y, and z axes) and $m$ is the atomic mass (for simplicity we start with equal masses which corresponds to, e.g., diamond and silicon particles). The eigenvectors of $D$ encode the displacements of all atoms in a phonon mode and the eigenvalues give the phonon frequencies.

Hamiltonian~\eqref{eq_hgen0} includes the details on valence angle bending term and thus can be derived based on, e.g., the Keating model~\cite{keating1966effect,anastassakis1990piezo}. When using the harmonic approximation and the Keating model for infinite crystals with diamond-type lattice one obtains the following approximate formula for the dispersion of optical phonons near the Brillouin zone center
\begin{equation}
\label{eq_dispersion}
\omega(q)=A+B\cos(\tilde{q} a_0/2)
\end{equation}
with certain parameters $A$ and $B$. This form of phonon dispersion assumes the following wave vector normalization\cite{ourDMM}: the maximal value of wave vector corresponding to the boundary of Brillouin zone is $\tilde{q}_{\rm max} \approx 2\pi/a_0$, where $a_0$ is the lattice parameter. In our calculations, we take a diamond with $a_0 \approx 0.357$~nm  as a representative and important for practice example and use 
\be
  A = 1266.22 \, \mathrm{cm}^{-1}, \, B= 66.8 \, \mathrm{cm}^{-1},
\ee
which yield
\be
  \omega_0 = 1333 \, \mathrm{cm}^{-1}.
\ee
For theoretical discussion, we shall use the redefined dimensionless momentum, $q = \tilde{q} a_0 /2 $ whose maximal value is $\pi$.

For optical phonons in nanoparticles, DMM demonstrates the size quantization effect~\cite{meilakhs2017new,ourDMM} and validity of scalar continuous Klein-Fock-Gordon equation within the Euclidean metric~\cite{ourEKFG} (EKFG), which can be used to derive the envelope functions $Y$ for atomic displacements corresponding to the optical modes. Explicitly the envelopes are given by
\begin{equation} \label{Laplace}
  \Delta Y + q^2 Y=0,  \qquad Y|_{\partial \Omega} = 0.
\end{equation}
and the spectrum of optical phonons valid both for infinite space ($q$ is the wave vector of the plane wave) and in the confined manifold ($q^2$ gives the size- and shape-dependent eigenvalue) reads
\begin{equation}\label{Spec3}
  \omega^2=C_2-C_1 q^2,
\end{equation}
which matches with Eq.~\eqref{eq_dispersion} in the long-wavelength limit.

After derivation of phonon modes labeled by index $\nu$, we employ the Bond polarization model~\cite{go1975bond,snoke1993bond} (BPM) to calculate their contributions into polarizability of a given vibrational mode and thus to the total polarizability of a nanoparticle. Then one directly reconstructs the Raman spectrum (RS) of the nanoparticle using the phonon frequencies and corresponding polarizability tensors
\begin{equation}
\label{eq_ramanvianormal}
I(\omega)\propto \sum_\nu \frac{n(\omega_{\nu})+1}{\omega_{\nu}}  \hat{N}(P_{\alpha\beta}(\nu)) \frac{\Gamma_\nu/2}{(\omega-\omega_\nu)^2+\Gamma_\nu^2/4},
\end{equation}
where $n(\omega_\nu)$ is the Bose-Einstein filling factor at a given temperature, mode intensity $\hat{N}(P_{\alpha\beta}(\nu))$ depends on incident and scattered light polarizations, nanoparticle orientation, and particular phonon mode $\nu$. The parameter $\Gamma_\nu$ represents the (mode-dependent) broadening. Various mechanisms of its origin were discussed in Refs.~\cite{our3,our4,ourSurf} in detail. In the final step, the Raman spectra of individual particles~\eqref{eq_ramanvianormal} should be summed up according to the size-distribution function of a powder. In the scalar EKFG model, one can obtain a simple equation for $\hat{N}(P_{\alpha\beta}(\nu))$. The polarizations are irrelevant in this case, and we simply have~\cite{ourEKFG}
\be
  \hat{N}_\nu \propto \left| \int Y_\nu d\mathbf{r}\right|^2.
\ee
For instance, in spherical particles, $Y$ are conventional spherical waves and the mode with the dominating contribution to RS is the one with $n=1, l=m=0$. Explicitly,
\be
  Y_1(\m{r}) = \frac{\sin{(2 \pi r/L)}}{2 \pi r/L}.
\ee
where $L$ is the particle size (diameter of the sphere).

Importantly, DMM can be naturally used to obtain the vibrational modes of nanoparticles with a disorder, which can be represented by, e.g., impurity atoms and/or defect bonds. Then, the summation of individual RS in the ensemble will be equivalent to the averaging over disorder configurations, and well-known in condensed matter physics tricks can be used.



\subsection{Theoretical concepts}
\label{SThC}

Due to the size quantization effect, the phonon lines \textit{broadening} in nanoparticles can follow two qualitatively different scenarios, namely, separated and overlapped regimes~\cite{our3,our4}. The former is characterized by well-resolved isolated peaks in the particle ensemble density of states (DOS) which are slightly broadened due to weak disorder. The latter is realized at moderate disorder strengths, when the phonon DOS resembles the bulk one, and the contributions of isolated (due to size-quantization in each particle) levels are hardly distinguishable.

In more detail, assuming weak impurities in the form of atoms with masses that differ from the masses of atoms in the parental compound, we write the disorder strength as
\be
  \label{eq_Sdefinition}
  S = \frac{\langle \delta m^2_l \rangle}{m^2} \approx c_{\rm imp} \left( \frac{\delta m}{m}\right)^2,
\ee
where $l$ enumerates the lattice sites, $\langle ... \rangle$ is averaging over all sites, and the last approximate equation refers to the case of weak binary disorder where we have a fraction with $c_{\rm imp} \ll 1$ of atoms with the mass $M = m+\delta m$ and $|\delta m| \ll m$. Analysis shows that for 4~nm-sized diamond particles, the boundary between the regimes of separated and overlapped levels is $S \sim 0.005$. Importantly, for separated levels, one has $\Gamma \propto \sqrt{S}$, whereas for overlapped ones conventional dependence $\Gamma \propto S$ reveals.

Importantly, from the experimental point of view, since the broadening at the crossover between the regimes is $\sim 1$~cm$^{-1}$, the separated regime observation is a challenging problem. So, in practice one can consider exclusively overlapped levels regime to deal with actual experimental data and further theoretical considerations will be focused on this limit. If in estimations one obtains the values for shift and broadening lower than 1~cm$^{-1}$, one can just neglect them.

Noteworthy, in the case of separated levels, mesoscopic fluctuations of the particle composition (i.e. fluctuation of the mean atomic mass among the nanoparticles in the ensemble) play an important role in providing an additional contribution to the broadening $\propto \sqrt{S}$. 

In contrast, when the disorder-induced \textit{shift} of the main Raman peak is addressed, the basic understanding comes from the fact that large phonon momenta provide the major contribution to this effect. Then, the size quantization is less important in this case and one can expect similar results in both regimes of separated and overlapped levels.

We proceed with the binary disorder. Analytical treatment is based on the simplified spectrum [cf. Eq.~\eqref{eq_dispersion}]
\be \label{spec1}
  \omega_{\m{q}} = \omega_0 - \alpha q^2.
\ee
For the diamond optical phonon dispersion~\eqref{eq_dispersion}, we have   $\alpha = B/2 = 33.4 $~cm$^{-1}$.

Near the positive frequency pole, one obtains the bare phonon Green's function in the form (cf. Ref.~\cite{our3}):
\be
  D_0(\omega,\m{q}) = \frac{1}{\omega - \omega_\m{q} + i 0},
\ee
which allows us to map the problem at hand onto the previously developed theory of Bose (triplon) excitations in a quantum magnet with the diagonal disorder~\cite{yashenkin2016,utesov2021self}. In our case, the disorder strength parameter is
\be \label{DisParam}
  u = \frac{1}{4} \left( \omega_0 \frac{\delta m}{m + \delta m} - \frac{z \, \delta K}{m \omega_0} \right),
\ee
where we also allow the disorder in the rigidity of the springs connecting the impurity atoms with the regular ones and $z$ is the number of nearest neighbors. This model should correspond to real particles. Below, for simplicity, we shall mostly concentrate on the case of disorder in masses only, thus using
\be
  u = \frac{\omega_0}{4}  \frac{\delta m}{m + \delta m}
\ee
as the phonon ``on-site energy correction'' (diagonal disorder). Note that there are certain differences with the results of Ref.~\cite{utesov2021self} due to the dispersion with negative effective mass~\eqref{spec1} in the present study.

Here, it is necessary to introduce one crucial for the theory parameter $b(u)$ with the physical meaning of the inverse scattering length. Explicitly,
\be
  b(u) = k^\prime_D + \frac{4\pi \alpha}{u},
\ee
where $k^\prime_D = 2 k_D/ \pi \sim 1$ is a parameter related to the ``Debye'' momentum,  which plays the role of the reciprocal space cut-off in the theory. In case of $\delta m \ll m$, one can write 
\be
  b^{-1} \approx \omega_0 \left( \frac{\delta m }{16\pi\alpha m} \right) - \omega_0(k'_D \omega_0 + 16\pi\alpha) \left( \frac{\delta m }{16\pi\alpha m} \right)^2. \nn \\
\ee

The provided preliminary analysis shows that in the overlapped levels regime one can use the results of Ref.~\cite{utesov2021self} obtained in the framework of a self-consistent T-matrix approach (SCTMA) designed for bulk materials. In the case of non-resonant scattering, one has
\be \label{dw1}
  \delta \omega = - c_{\rm imp} \frac{4 \pi \alpha}{b(u)}
\ee
for the peak frequency correction and
\be \label{gamma1}
  \Gamma(\m{q}) = c_{\rm imp} q \frac{8 \pi \alpha}{b^2(u)}
\ee
for the damping of a quasiparticle with momentum $\m{q}$. These equations account for all orders of perturbation theory for the scattering off a single impurity. Noteworthy, for weak impurities Eq.~\eqref{dw1} includes all dominant contributions, i.e., the peak shift due to alteration of mean mass for binary disorder and the peak shift in the second-order perturbation theory for the Gaussian disorder. By combining Eqs.~\eqref{dw1} and~\eqref{gamma1}, we obtain the following important result:
\be \label{BinRel}
  \frac{\delta \omega}{\Gamma(\m{q})} = - \frac{b(u)}{2 q}.
\ee

In the popular for theoretical considerations Gaussian case, the first order in $b^{-1}$ contribution to $\delta \omega$ vanishes due to zero mean value of disorder, and the corresponding relation reads
\be \label{GaussRel}
  \frac{\delta \omega}{\Gamma(\m{q})} \approx \frac{k^\prime_D}{2 q}.
\ee

Importantly, we expect that the line broadening parameter~\eqref{gamma1} should also hold in confined geometry as for original consideration of plane waves in the bulk; the momentum $\m{q}$ should be taken as an effective wave vector of the standing wave in the confined geometry in this case. For the broadening of phonons which provide the main contribution to the Raman spectrum in the overlapped regime, one should use Eq.~\eqref{gamma1} with $q_L = \pi a_0/L$ approximately corresponding to the wave number of the first level of the size quantization in nanoparticle with the size $L$. Hence, for nanometer-sized particles, the ratio~\eqref{GaussRel} is $\sim 1$.


However, in the limit $|b| \to 0$ Eqs.~\eqref{dw1} and~\eqref{gamma1} are evidently not useful. It is a consequence of the \textit{resonant} scattering phenomenon that emerges due to the existence of a (quasi)localized state on impurity with the energy close to $\omega_0$. Explicitly, for $u < -4 \pi \alpha/k^\prime_D$ (light defect atom) in the single impurity problem, there exists a bound state with energy above the optical phonon band, which reads
\be
  \omega_{loc} = \omega_0 + \alpha b^2(u)
\ee
in the limit of $b(u) \ll k^\prime_D$.

In order to treat the problem at finite defect concentration even at $|b| \ll 1$ and obtain reasonable results, one can use, e.g., SCTMA~\cite{utesov2021self}. Its main yield is as follows. Eqs.~\eqref{dw1} and~\eqref{gamma1} are correct in the limit of $|b| \gtrsim (4 \pi c_{\rm imp})^{1/3}$. In the opposite case of $|b| \ll (4 \pi c_{\rm imp})^{1/3}$ one has
\be \label{dw2}
  \delta \omega \approx 3 \alpha (2 \pi c_{\rm imp})^{2/3} + 2^{2/3} \alpha b (2 \pi c_{\rm imp})^{1/3}
\ee
and
\be \label{gamma2}
  \Gamma(\m{q}) = \sqrt{3} \alpha  (2 \pi c_{\rm imp})^{2/3} + O(q^2)
\ee
for the frequency shift and damping, respectively. The last equation is written under an assumption that \mbox{$q \ll (4 \pi c_{\rm imp})^{1/3}$} which indicates that the quasiparticle wavelength is much larger than the average distance among impurities. Evidently, this is true for nanoparticles with many defect atoms since the phonon wavelength is of the order of particle size.

To summarize this section, we need to distinguish the three scenarios: (i) weak impurities for which $\delta \omega \propto c_{\rm imp}$ and $\Gamma \propto \sqrt{c_{\rm imp}}/L^{3/2}$, (ii) strong non-resonant impurities for which $\delta \omega \propto c_{\rm imp}$ and $\Gamma \propto c_{\rm imp}/L$, and (iii) strong resonant impurities for which $\delta \omega, \Gamma \propto c_{\rm imp}^{2/3}$. We observe all these regimes in our numerics. The corresponding results are discussed below.

\subsection {Raman peak shape and Green's function calculation details}
\label{SCalc}

Along with the analytical approach, the imaginary part of the phonon Green's function can be derived numerically from calculating the optical phonon modes of nanoparticles ensemble with many disorder realizations. In our numerics, we employ DMM~\cite{ourDMM} in spherical particles to address the contribution from lattice impurities to the shift of the Raman peak and compare it with the corresponding broadening. To construct the dynamical matrix we use the Keating model. In particular, we treat the Gaussian disorder in atomic masses, binary disorder at small concentrations, and vacancies.


We employ from 64 to 256 realizations depending on the considered disorder parameters and account for 1024 highest vibrational modes, which fully cover the actual frequency range above 1200~cm$^{-1}$ for 3~nm diamond particle. Next, we calculate the spectral weight (imaginary part of the Green's function) for $\nu$-th mode of the pure particle as follows:
\begin{equation}\label{eq_ImGreen1}
 \text{Im} \, D_\nu(\omega) = - \pi \sum_{i, \varepsilon} \delta(\omega- \omega_{i,\varepsilon}) | \langle \nu | i, \varepsilon \rangle|^2,
\end{equation}
where the summation spans all disorder realizations $i$ and the corresponding eigenstates $\varepsilon$ within each realization of the disorder. Finally, this spectral weight should be approximated by the Lorenz function in order to obtain the broadening $\Gamma_\nu$ and the energy shift with respect to the pure particle frequency $\omega_\nu$. The example of eigenfunction $\psi_{i,\varepsilon}({\bf r})$ in the presence of the disorder is given in Fig. 15 from Ref. \cite{our4} for the case of one single impurity in the nanoparticle.

We concentrate on the highest energy modes of the pure particle, which modification in the presence of disorder results in the RS shift and broadening. Their spatial structure (cf. Fig. 3 from Ref. \cite{ourDMM}) is the most symmetrical one similar to the first eigenmode of the EKFG approach $Y_1$. Thus, the matrix element
\begin{equation}\label{eq_greenexplicit}
    | \langle \nu | i, \varepsilon \rangle|^2 \equiv \left| \int \psi_{\nu}({\bf r}) \psi_{i,\varepsilon}({\bf r}) d{\bf r} \right|^2
\end{equation}
correlates with the Raman intensity of each particular mode
\begin{equation}\label{eq_structure_factor}
    I_{i, \varepsilon} (\omega) \propto \delta (\omega - \omega_{i,\varepsilon}) \cdot \left| \int \psi_{i,\varepsilon}({\bf r}) d{\bf r} \right|^2.
\end{equation}
Hence, the main Raman peak can be well approximated by the corresponding spectral weight. As it is shown below, it was indeed observed in our numerics.

\section{Results}
\label{SRes}

\subsection{Gaussian disorder}
\label{SGauss}

\begin{figure}
  \includegraphics[width=8.cm]{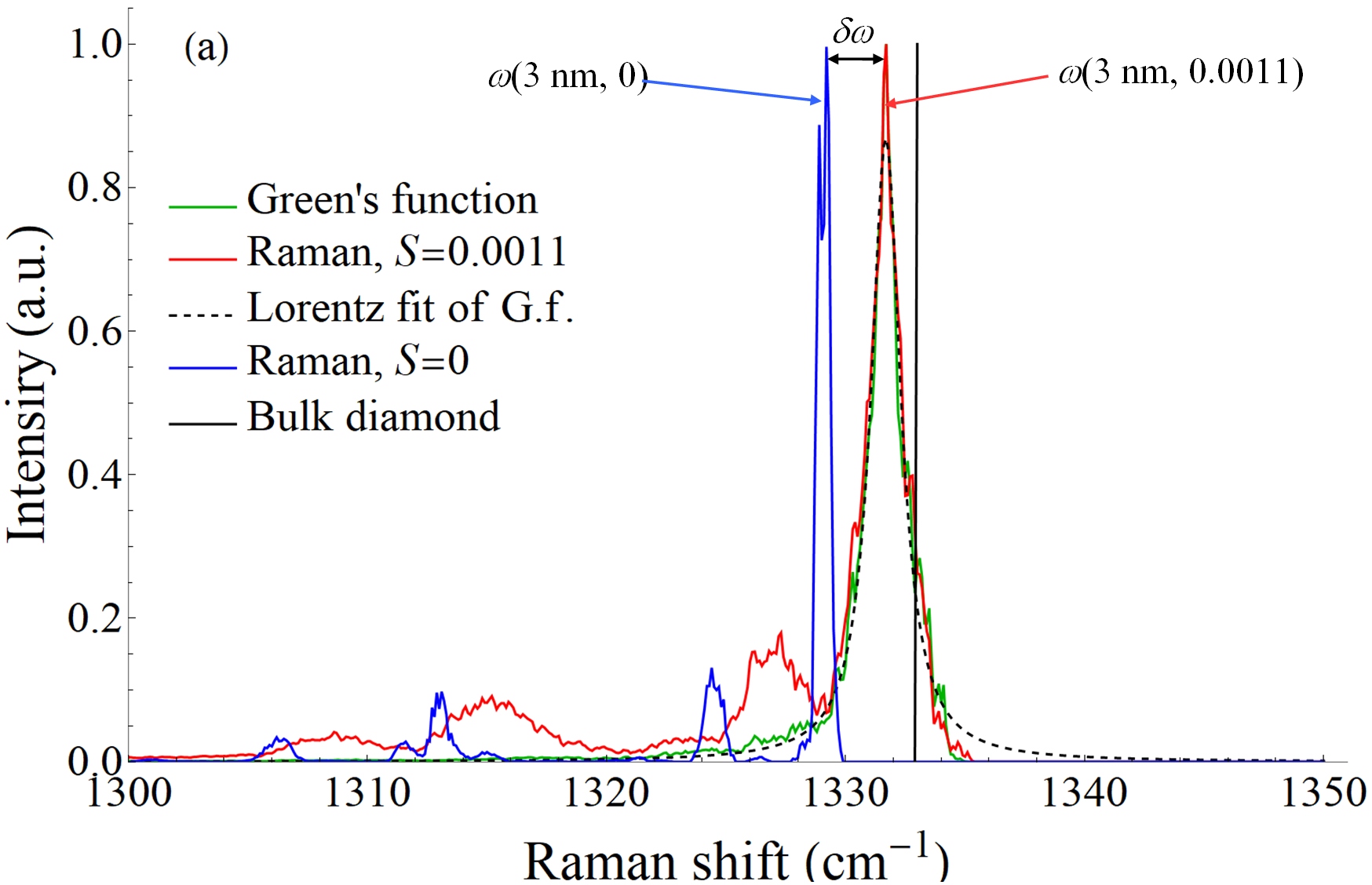} \\
  \includegraphics[width=8.cm]{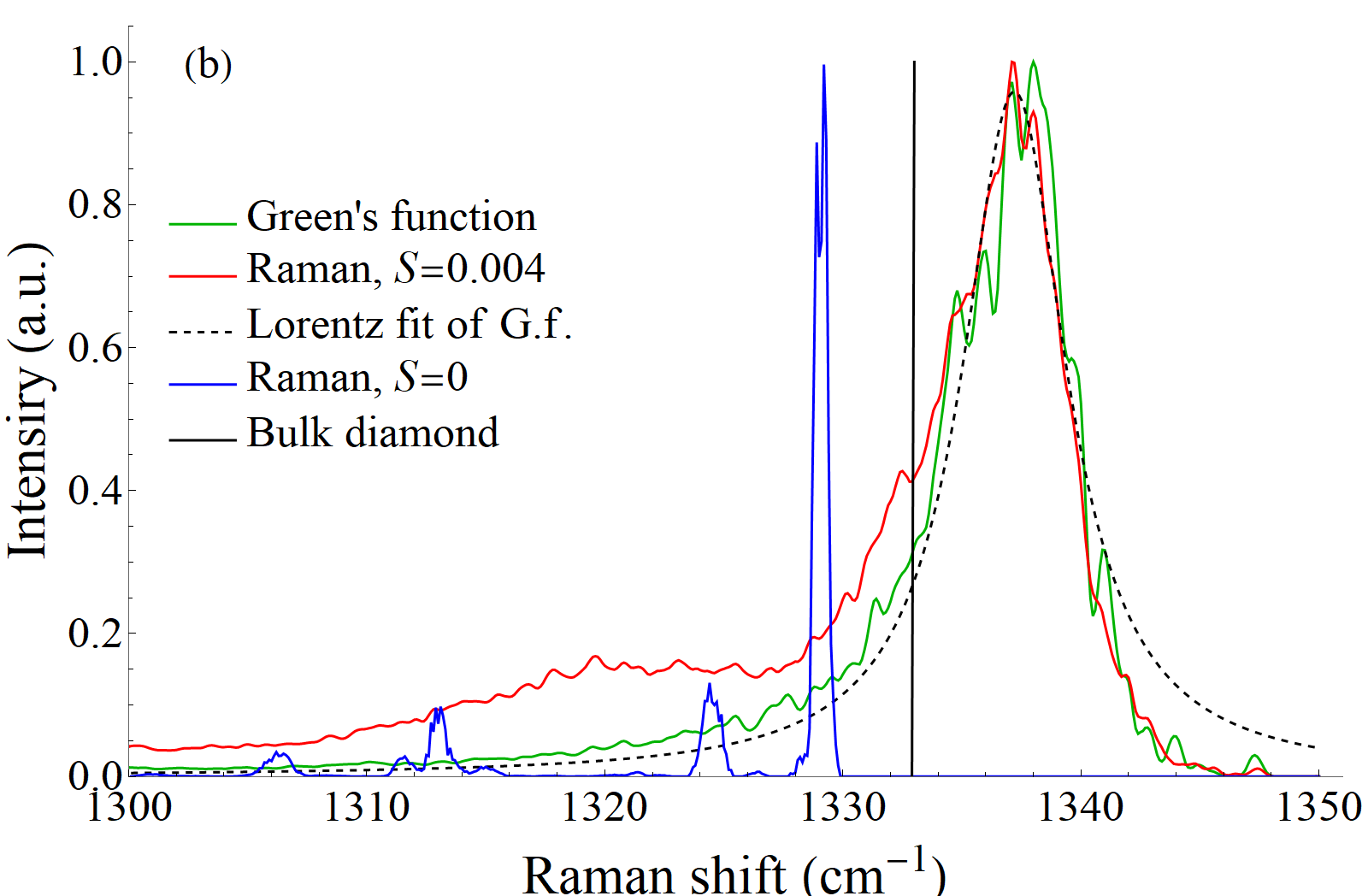}
  \caption{Raman spectra (red curves) and highest phonon mode Green's function imaginary part (green curves) for various Gaussian disorder strengths obtained numerically for the ensemble of 3 nm diamond particles. (a) $S=0.0011$ (weak disorder case) and (b) $S=0.004$ (moderate disorder case). The black dashed curve stands for the spectral weight peak fit with the Lorentzian curve. The blue curve shows the slightly broadened by hands Raman spectrum of pure 3~nm particle and the vertical black line shows the energy of $q=0$ phonon in bulk diamond $\omega_0=1333$~cm$^{-1}$. Note that in both panels RS is blueshifted due to the disorder.}
\label{fig_1_examplesGFspectra}
\end{figure}

We start with the simplest and probably the most popular model of the disorder, which can be characterized by a single parameter measuring its strength $S$. Fig.~\ref{fig_1_examplesGFspectra} illustrates the claim that the Green's function imaginary part (proportional to the spectral weight) for the highest symmetrical mode and the Raman spectrum should be very similar. Here we consider a spherical diamond particle of size $L=3$~nm subject to the Gaussian disorder in masses and the spectrum averaged over 256 realizations. Note that the discrepancy between the RS and the spectral weight can be observed on the left shoulder of the peak. Its origin is due to the existence of Raman-active modes with lower energy. However, their contribution is significantly smaller in comparison with the highest in energy mode~\cite{ourDMM}.

Since the main goal of this study is the analysis of the relation between RS broadening and shift, it is necessary to define the disorder-induced Raman peak shift explicitly. In the Gaussian case, we shall use notation
\be
  \delta \omega(L,S) = \omega(L,S) - \omega(L,0)
\ee
for the peak shift in the presence of the disorder with the strengths $S$ for particles of the size $L$. Note that $\omega(L \to \infty,0) = \omega_0$ corresponds to a pure bulk diamond which can be calculated using, e.g., \textit{ab initio} methods. Importantly, with the high accuracy $\omega(L,0)$ is given by Eq.~\eqref{spec1} for $q = 2 \pi a_0/L$.  For 3~nm diamond particles $\omega(L,0) = 1328.8$~cm$^{-1}$. In Fig.~\ref{fig_1_examplesGFspectra} one can see that the Gaussian disorder shifts the peak to larger frequencies and this effect is pronounced even at relatively small $S$. Panel (a) corresponds to the separated levels regime (peaks from various modes do not touch each other) and panel (b) is for the overlapped one with indistinguishable peaks.

\begin{figure}
  \centering
  \includegraphics[width=8.cm]{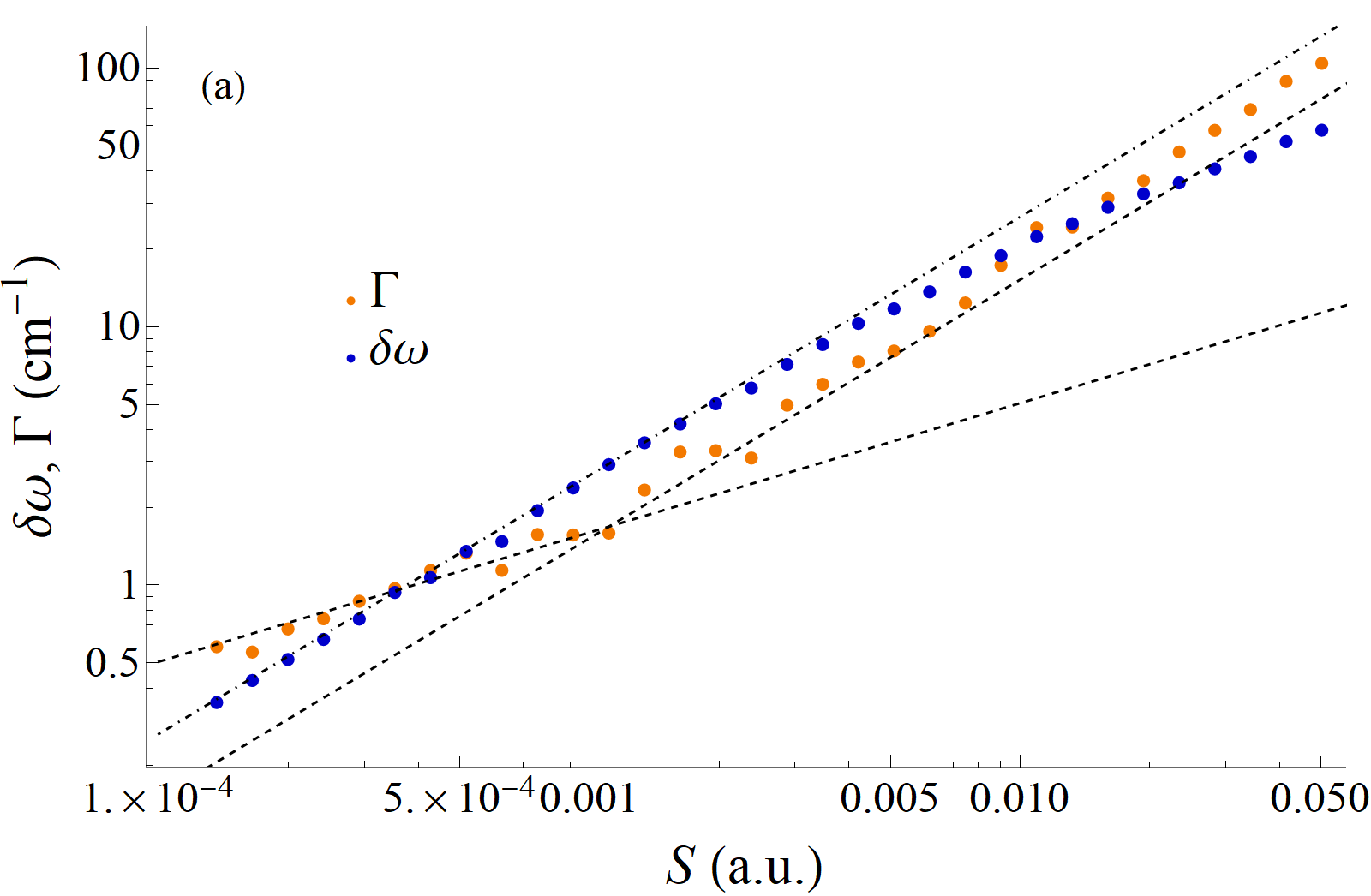} \\
  \includegraphics[width=8.cm]{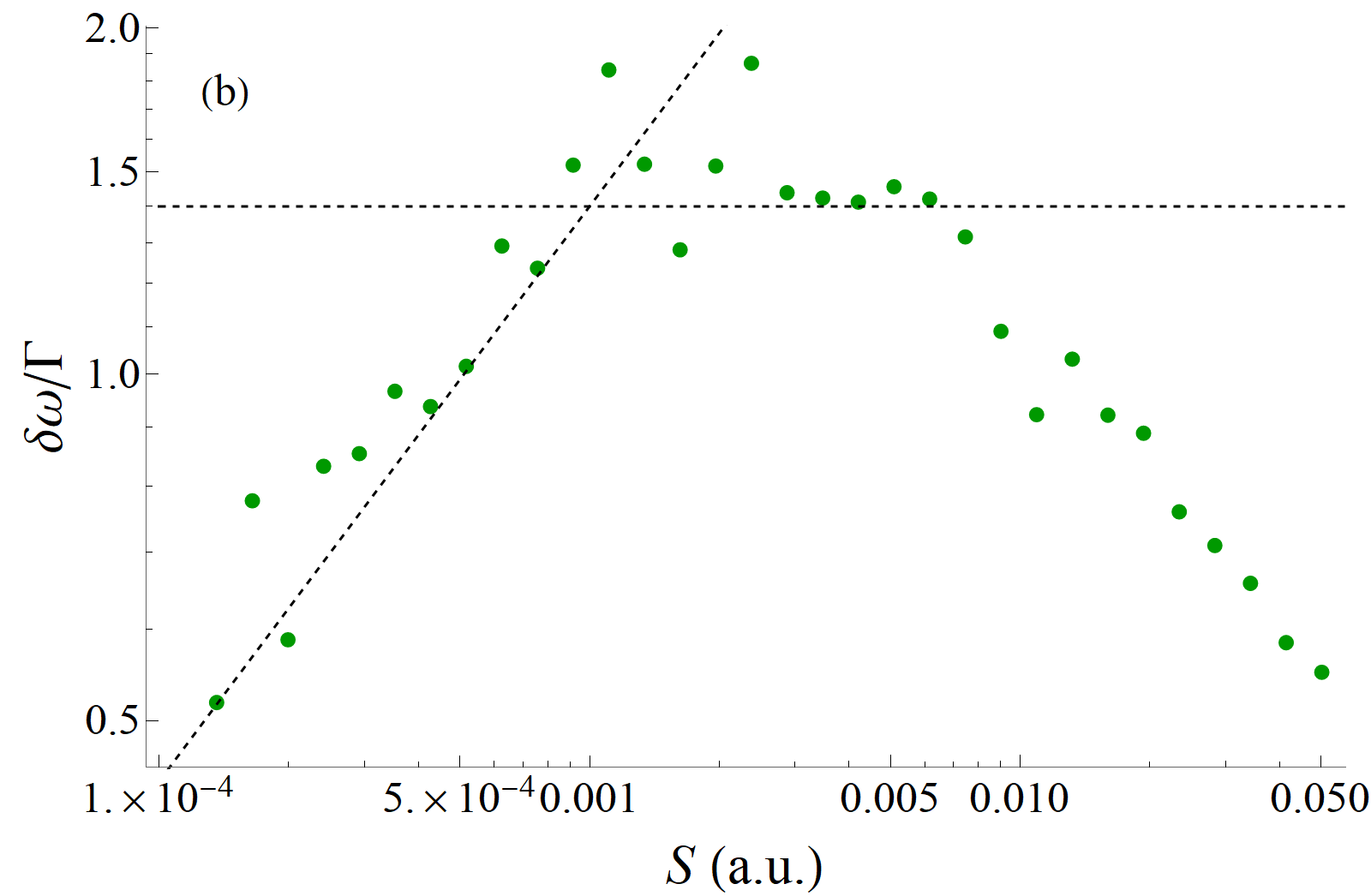}
  \caption{(a) Broadening $\Gamma$ and the peak blueshift in ensemble of 3~nm diamond particles as a function of Gaussian disorder strength $S$. The results are obtained from the Lorentzian fit of curves calculated within DMM. Eye guides for $\sqrt{S}$ and linear function are given. (b) The ratio of the blueshift and broadening as a function of the disorder strength.}
\label{fig_2_gaussianGammaDW}
\end{figure}

\begin{figure}
  \centering
  \includegraphics[width=8.cm]{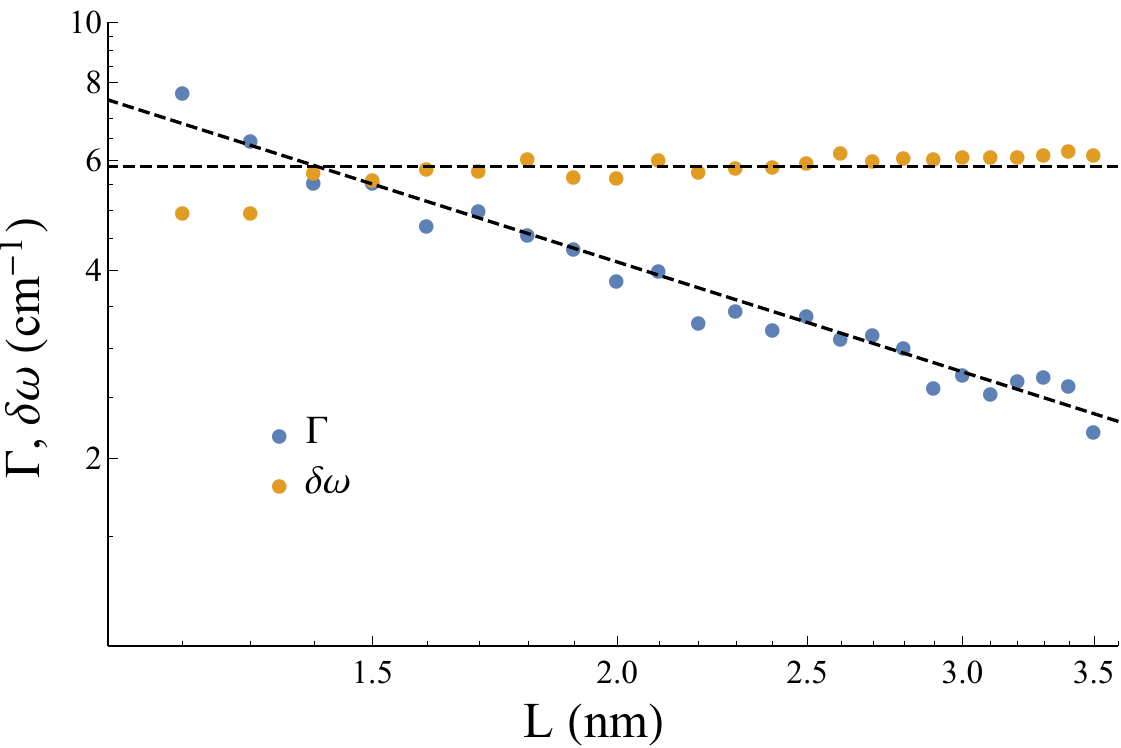} \\
  \caption{In the case of moderate disorder (i.e., overlapped levels regime, see text) the peak shift is almost independent of the particle size, whereas the broadening is $\propto 1/L$. Here the Gaussian disorder with $S=0.0018$ is considered. Dashed lines are best fits of $\Gamma$ and $\delta \omega$ with $1/L$ and $const$ functions, respectively.  }
\label{fig_3_GaussL}
\end{figure}

Next, for the same ensemble of nanodiamonds, we study the peak shift $\delta \omega$ and its broadening $\Gamma$ as functions of $S$, see Fig.~\ref{fig_2_gaussianGammaDW}. For the broadening, the two regimes are visible, which agrees with analytical predictions. For weak disorder, the separated levels regime takes place. The levels are slightly broadened with $\Gamma \propto \sqrt{S}$ near the unperturbed positions and their spectral weights do not overlap. For the overlapped levels regime, the disorder leads to touching and overlap of the broadened peaks, and the conventional law $\Gamma \propto S$ is observed. The crossover between these regimes takes place at $S \sim 0.001$ for considered here 3~nm diamond particles. It corresponds to $\Gamma \sim 1$~cm$^{-1}$. This ``critical'' $\Gamma$ and $S$ were shown to depend only on the nanoparticle size and phonon dispersion~\cite{our3}. Note that in both regimes, the shift behavior is the same, $\delta \omega \propto S$.  However, this dependence is evidently violated when $S \gtrsim 0.01$ [see Fig.~\ref{fig_2_gaussianGammaDW}(a)] what can be understood as follows: for strong Gaussian disorder some atoms are significantly lighter than the average atom of a particle and they lead to the appearance of localized vibrations which push propagating phonon lines toward lower frequencies (see also discussion below). Hence, for strong Gaussian disorder, $\delta \omega$ shows a tendency to saturation in contrast to $\Gamma$. However, in the wide range of moderate disorder strengths, we observe [see Fig.~\ref{fig_2_gaussianGammaDW}(b)] that $\delta \omega \propto \Gamma$ with the coefficient being $\approx 1.4$. So [cf. Eq.~\eqref{GaussRel}] for particles of the size $L$~nm one should have 
\be
  \delta \omega = 1.4 \cdot \frac{L}{3~\textrm{nm}} \cdot \Gamma.
\ee
This formula is also supported by the results shown in Fig.~\ref{fig_3_GaussL}. 

An intuitive explanation of the characteristic for the Gaussian disorder blueshift [$\delta \omega (L,S) > 0$] can be formulated as follows: wave-function of the highest frequency state tends to occupy regions where there are more light atoms and avoid regions where heavier atoms dominate. On the technical level, we note that the first-order corrections to frequencies are zero due to the Gaussian type of disorder with zero average. However, the second-order correction is always positive~\cite{landau2013quantum} because we are focusing on the state with the highest frequency. Thus, we conclude that the Gaussian disorder provides the shift of the main Raman peak in the opposite direction to the size-quantization effect (or equivalently, the phonon confinement) which is responsible for the redshift of the peak with respect to its bulk position.

\subsection{Binary disorder}

The Gaussian disorder is a convenient model to study since it is characterized by a single parameter $S$. However, in real materials, there are usually certain types of point defects. For example, in diamond particles, it can be $^{13}$C isotopes or nitrogen-vacancy complexes (so-called NV centers). For their description, the binary disorder model is more suitable. The peak shift and broadening in this case are dependent on the particle size, concentration, and the type of defect (e.g., the mass of the impurity atom).

We start the presentation of our results with the case of binary disorder in atomic masses. Note that to make a comparison with the theoretical predictions (see Subsec.~\ref{SThC}) more precise, we shall also discuss the unphysical impurity masses, i.e., not existing in nature. Those who want to employ our results for the description of real impurities can simply ignore the unphysical cases. 

\begin{figure}
  \centering
  \includegraphics[width=8.cm]{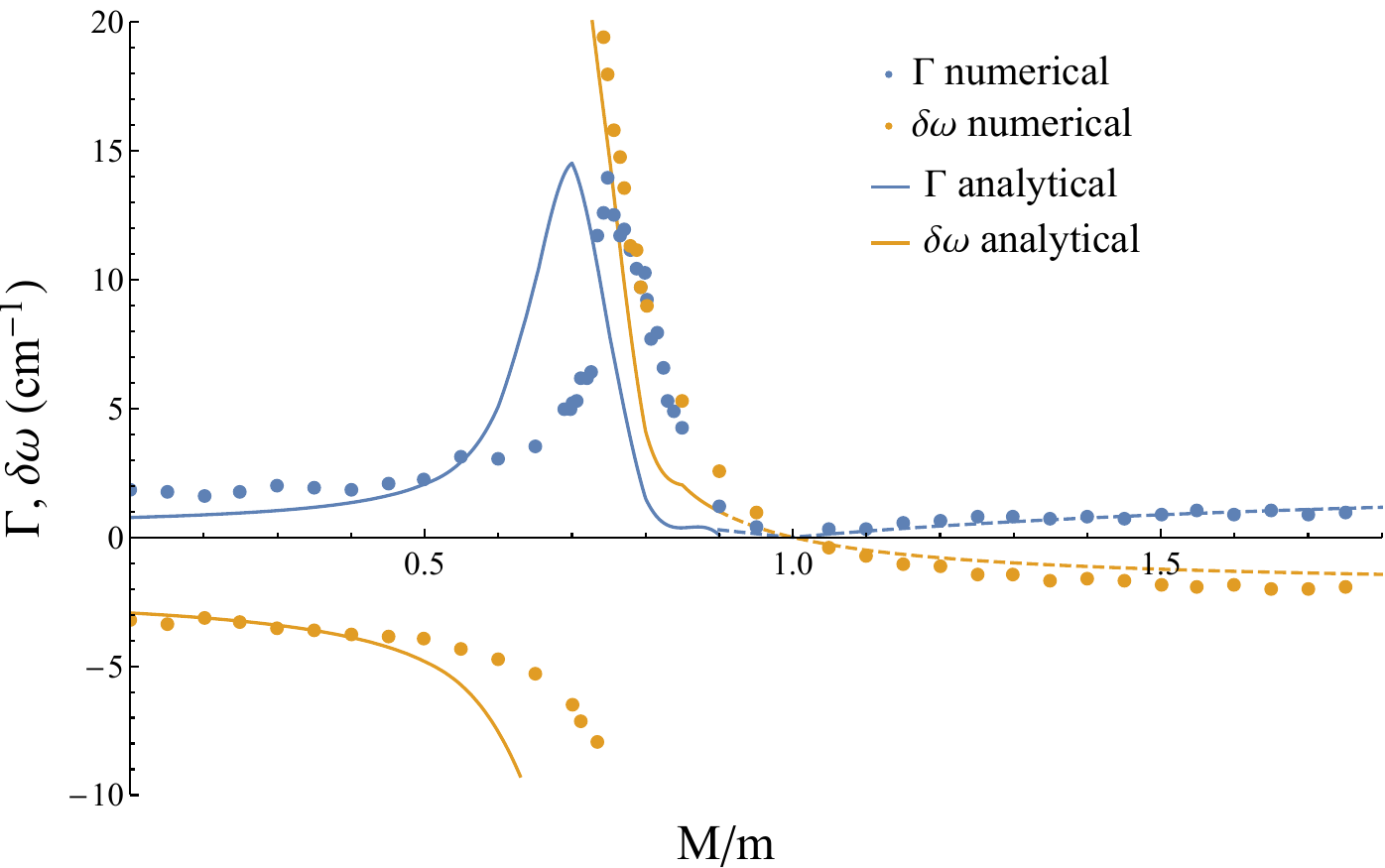} \\
  \caption{Overview of the numerical and analytical results (see text) for various impurity masses $M$ at fixed $c_{\rm imp} = 0.02$. The points for broadening $\Gamma$ and the peak shift $\delta \omega$ were obtained numerically for 3~nm diamond particles. Solid and dashed lines correspond to analytical estimations for the overlapped and the separated regimes, respectively.  }
\label{fig_4_masses}
\end{figure}

Fig.~\ref{fig_4_masses} summarizes our results for impurity masses covering a broad range from light to heavy $M$. For 3~nm diamond particles and impurity concentration $c_{\rm imp}=0.02$, we observe, roughly speaking, three qualitatively different regimes: (i) separated levels regime for $M/m > 0.9$ (weak and heavy impurities), (ii) non-resonant overlapped regime, and (iii) resonant overlapped regime.  We discuss them in detail below using theoretical concepts of Subsec.~\ref{SThC}.

The first one is characterized by weak broadening and small negative $\delta \omega$ (redshift). The results are well fitted by the analytical prediction
\be
  \Gamma \propto \sqrt{c_{\rm imp} u^2},
\ee
and $\delta \omega$ is well approximated by Eq.~\eqref{dw1}. For the parameters above we have $|\delta \omega|/\Gamma \approx 1.8$.

\begin{figure}
  \centering
  \includegraphics[width=8.cm]{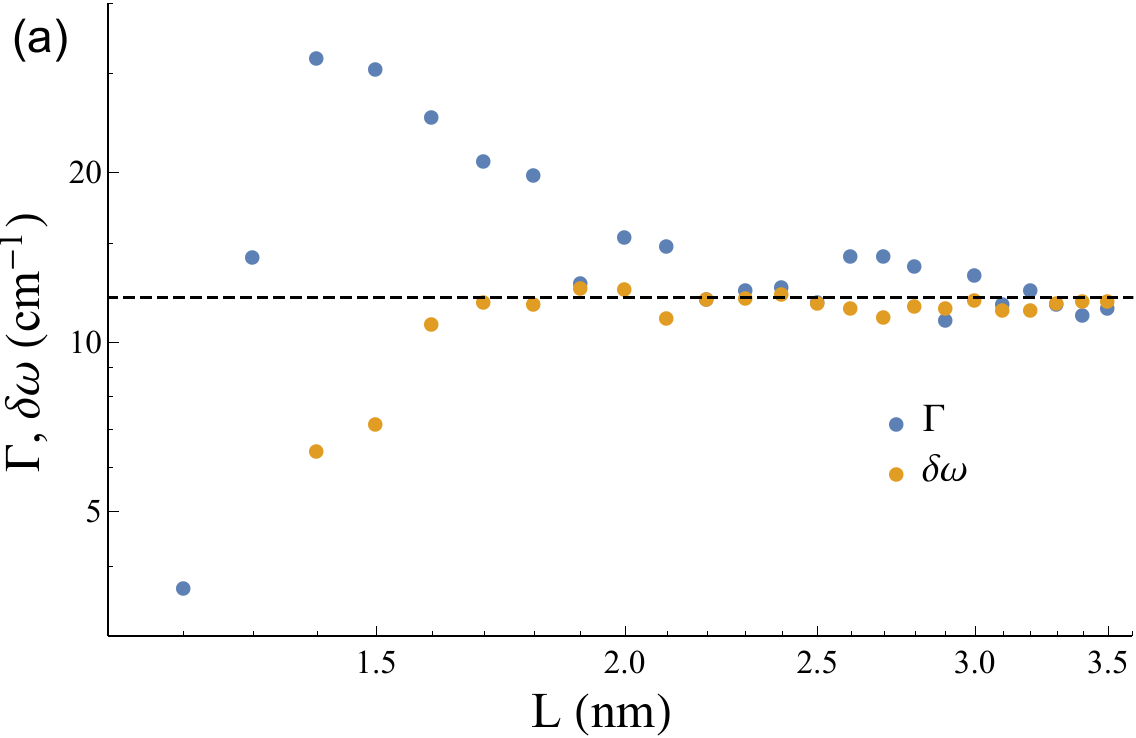} \\
  \includegraphics[width=8.cm]{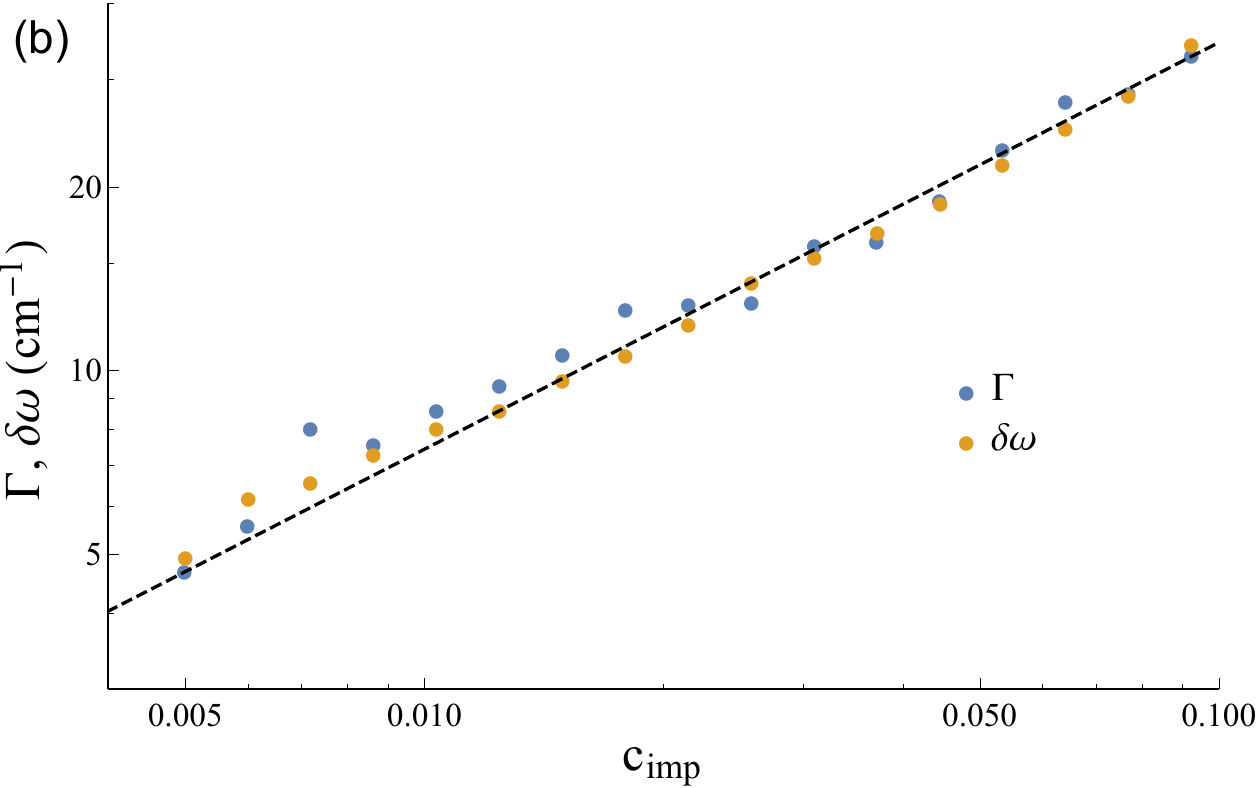} \\
  \caption{(a) In the case of (nearly) resonant scattering, the peak shift $\delta \omega$ and its broadening $\Gamma$ are almost independent of particle size for not very small particles. The dashed line is a guide for an eye. (b) In the resonant regime, both $\Gamma$ and $\delta \omega$ are $\propto c^{2/3}_{\rm imp}$ in a wide range of impurity concentrations. Here the dashed line is a guide for an eye indicating power-law with exponent $2/3$.}
\label{fig_5_res}
\end{figure}

The crucial feature of the overlapped levels regime ($M/m <0.9$ for chosen parameters) is the appearance of resonant scattering. For our particular parameters, the numerical analysis shows that the threshold for the bound state localized on the impurity to exist is $M/m \approx 0.78$. Analytically, we can assess the region of resonant scattering using the criterion $|b(u)| \sim (4 \pi c_{\rm imp})^{1/3} $, which yields an estimate $0.73 < M/m < 0.83 $. Outside this domain, the scattering can be treated as non-resonant.  It is not a surprise that the analytical equations presented in Sec.~\ref{SThC} could provide only a semi-quantitative description of the numerical data. The main reasons are that they were derived for a simple lattice with one atom in a unit cell under the assumption of quadratic dispersion in a spherical Brillouin zone, which also introduces the $k_D$ parameter. Nevertheless, our estimations with $k^\prime_D = 3$ show a reasonable agreement with the numerics. The main difference is the threshold value $M/m \approx 0.7$ which leads to the slightly different position of the resonant peak in $\Gamma$ (see Fig.~\ref{fig_4_masses}). Thus, Eq.~\eqref{BinRel} provides a good estimate of the relation between the shift and the broadening in the non-resonant regime, whereas for the resonant one Eqs.~\eqref{dw2} and~\eqref{gamma2} are useful. In particular, the latter yield $\delta \omega/ \Gamma$ independent of $L$ which was observed in our numerics for not very small particles, see Fig.~\ref{fig_5_res}(a). Note that the standard T-matrix approach yields $\Gamma \propto L$ in the resonant scattering case~\cite{our4}. Furthermore, our numerics also support the theoretically predicted $\Gamma, \, \delta \omega \propto c^{2/3}_{\rm imp}$ dependencies, see Fig.~\ref{fig_5_res}(b). The best fits of the data for $\Gamma$ and $\delta \omega$, give exponents $0.63$ and $0.67$, respectively. These values are very close to the analytically predicted $2/3$. 

\begin{figure}
  \centering
  \includegraphics[width=8.cm]{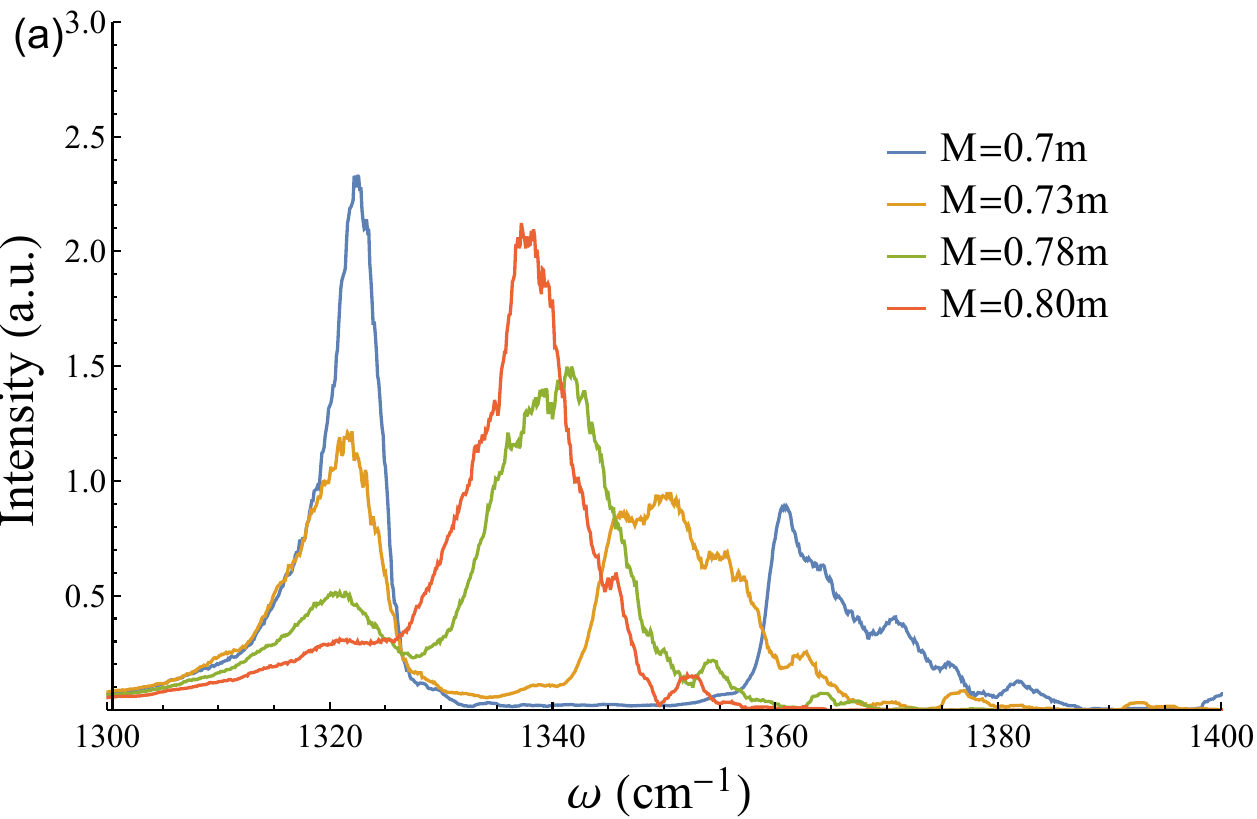} \\
  \includegraphics[width=8.cm]{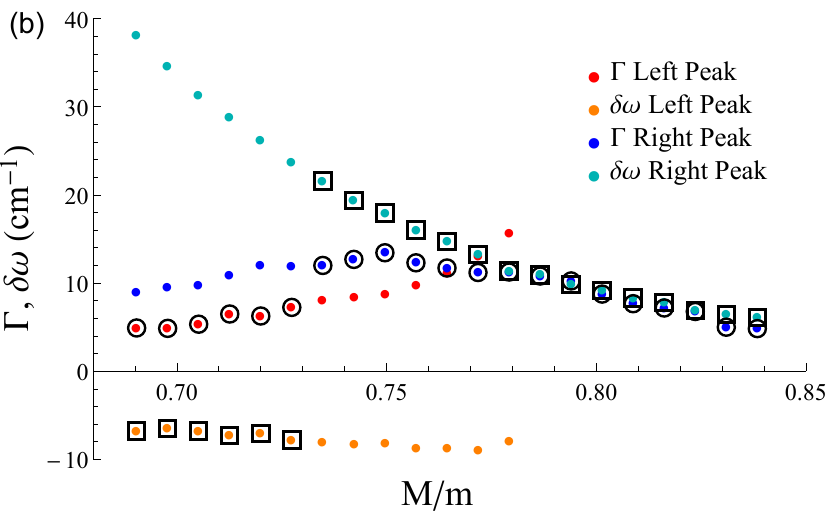} \\
  \caption{(a) Raman spectra of 3~nm diamond particles with impurities of various masses close to the resonant one. The data are smoothed for illustration purposes. (b) Dependence of broadening and shift for left (if exists) and right peaks on the impurity mass. Black circles (for $\Gamma$) and squares (for $\delta \omega$) indicate the properties of the peak providing the main contribution to the Raman spectrum (see text). In both panels  $c_{\rm imp}=0.02$.}
\label{fig_6_4masses}
\end{figure}

The properties of the Raman spectra in the resonant scattering regime deserve deeper discussion. We plot them for various $M/m$ in Fig.~\ref{fig_6_4masses}(a). For $M/m=0.8$ slightly above the localized state appearance threshold, there is a single broad peak. At the resonance $M/m=0.78$ the peak broadening and blueshift increase, and some satellite signal emerges. Upon further $M$ decrease, the satellite intensity grows up as well as its separation with the main peak. At $M/m =0.73$ their spectral weights are almost identical. In this case, the eigenmodes correspond to strongly hybridized propagating and localized on impurities states. Finally, for lighter non-resonant impurities the former satellite provides the main intensity accompanied by the ``band'' of localized states with frequencies substantially larger than $\omega_0$. Importantly, the latter pushes the main peak to the lower frequencies, which results in RS redshift (cf. Fig.~\ref{fig_4_masses}). A more detailed picture of this double-peak structure is discussed in Fig.~\ref{fig_6_4masses}(b). 
 
\begin{figure}[t]
  \centering
  \includegraphics[width=8.cm]{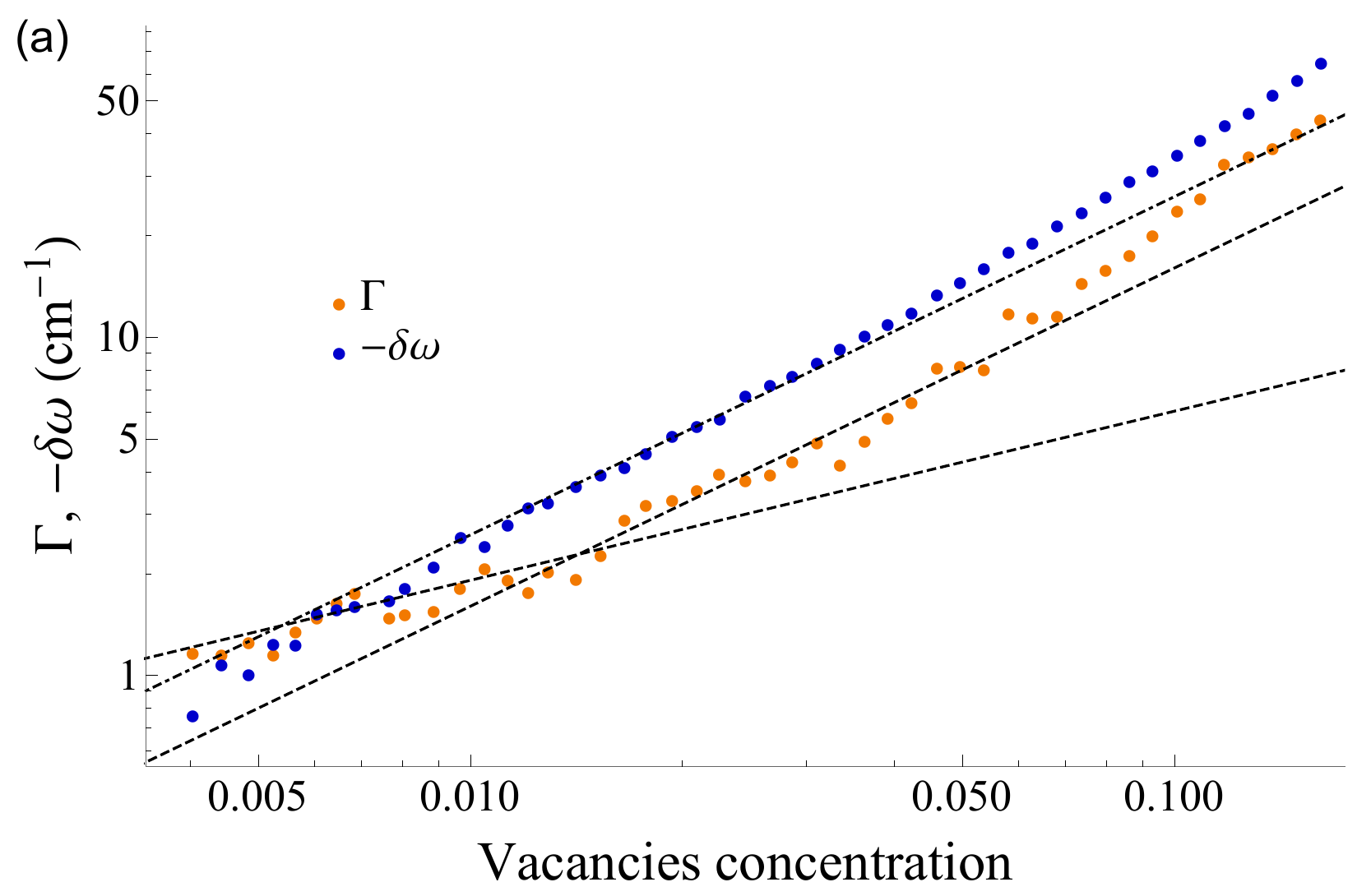} \\
  \includegraphics[width=8.cm]{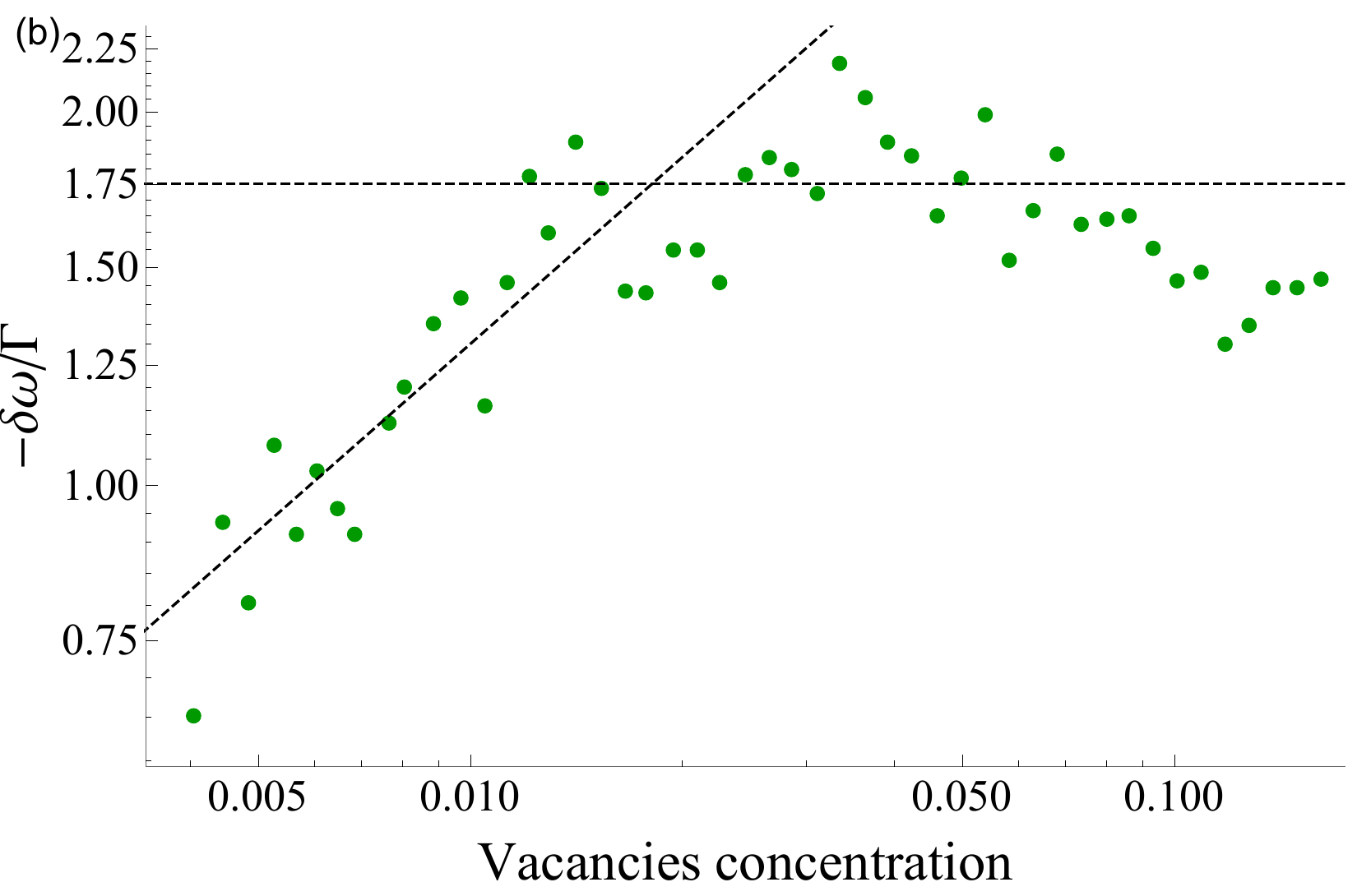}
  \caption{(a) Broadening $\Gamma$ and the peak shift in ensemble of 3~nm disordered nanodiamonds as a function of vacancies concentration. (b) The ratio of blueshift and broadening for various concentrations. In the broad range of concentrations, the ratio is approximately constant.}
\label{fig_vac}
\end{figure}

Now we turn to the particular point-like defect case, vacancies. Importantly, it is the main ingredient of the NV center since heavy atoms play a minor role in our consideration (see above). In our numerics, the vacancy is simulated as an absence of the atom in the diamond crystalline lattice. Out findings are illustrated in Figs.~\ref{fig_vac}(a) and~(b). The linear in impurities concentration overlapped levels regime starts from the vacancies concentration $c_{\rm vac} \approx 0.01$ and $\Gamma \approx 2$~cm$^{-1}$. At higher vacancy concentrations $c_{\rm vac} \gtrsim 0.1 $ the linear dependencies are evidently violated. Fig~\ref{fig_vac}(b) shows that in a broad range of concentrations, the ratio between $\delta \omega$ and $\Gamma$ is approximately constant being $\approx -1.75$ for 3~nm diamonds. Thus
\be \label{VacRel}
  \delta \omega = -1.75 \cdot \frac{L}{3~\textrm{nm}} \cdot \Gamma.
\ee
Analytically, the vacancies correspond to the so-called unitary limit $u \to -\infty$ for which
\be \label{BinRelUnit}
  \frac{\delta \omega}{\Gamma(\m{q})} = - \frac{k^\prime_D}{2 q}.
\ee
Note that this ratio is opposite to the Gaussian case [see Eq.~\eqref{GaussRel}].

Qualitatively the negative values of $\delta \omega$ can be explained by additional confinement of phonon wave function between the vacancies, which results in acquiring higher values of effective momentum and thus the lower energy of vibrations due to the dispersion with negative effective mass.

\begin{figure}[t]
  \centering
  \includegraphics[width=8.cm]{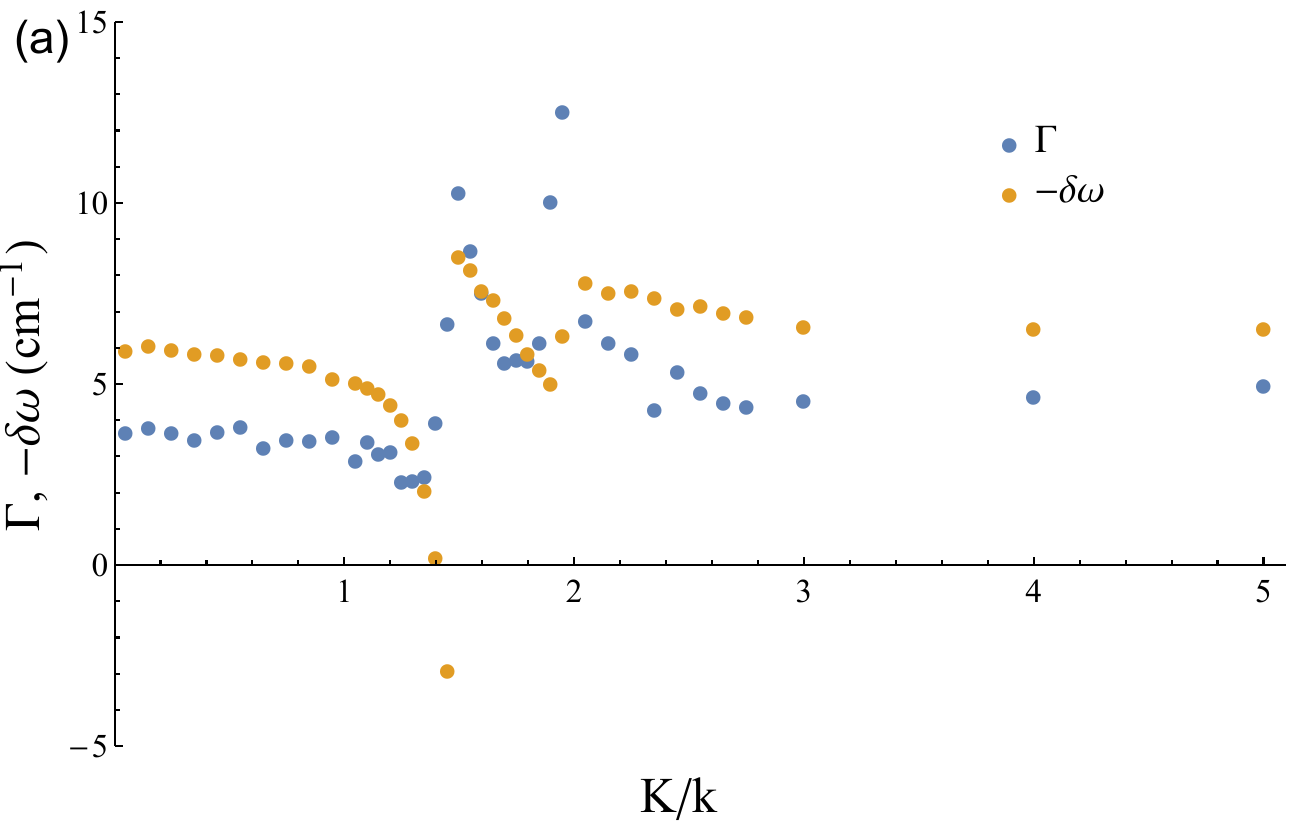} \\
   \includegraphics[width=8.cm]{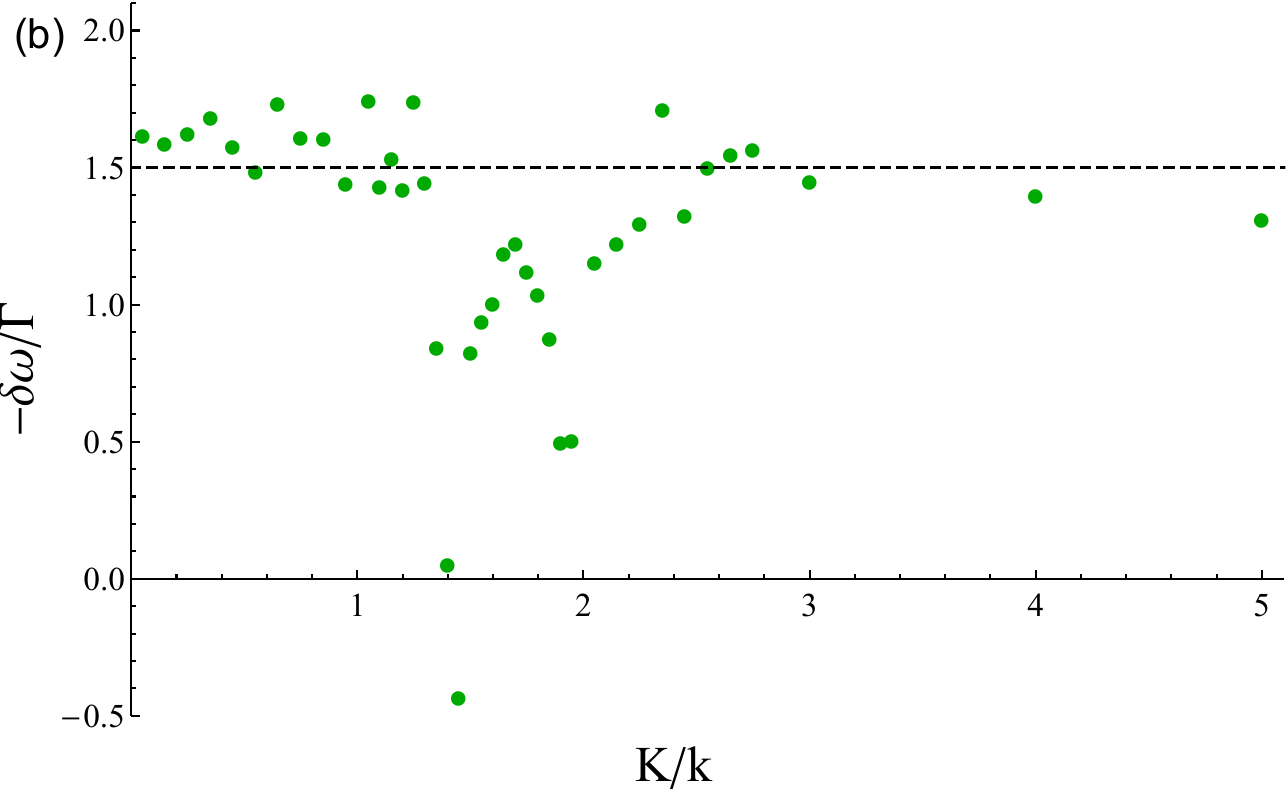} \\
  \caption{ (a) The broadening and the peak shift for 3~nm diamond particles with 2\% of NV centers as a function of N-C to C-C elastic constants ratio. (b) $-\delta \omega/\Gamma$ ratio for the same parameters. The dashed line is a guide for an eye.}
  \label{fig_NV}
\end{figure}

Next, we study NV centers which are typical for diamonds. To do so, we employ a simple model, where the center consists of a heavy atom (nitrogen), vacancy, and modified elastic constants for the corresponding N-C bonds. In the absence of knowledge about the latter, we scan a wide range of the ratios between elastic constants, see Figs.~\ref{fig_NV}(a) and (b), where the results for 3~nm diamonds are studied. Except for some resonant-like features located at $K/k \approx 1.5$ and $K/k \approx 1.95$, we observe that the broadening and the shift are weakly dependent on $K$. As in the case of vacancies, the NV centers contribute to the Raman peak redshift. Moreover, a typical value of $-\delta \omega/\Gamma$ is also similar being $\approx 1.5$. So, Eq.~\eqref{VacRel} should be useful in both cases.  Noteworthy, according to the literature, the covalent bond force constant essentially depends on the order of the bond, but not on the nature of atoms forming the bond. Following Refs. \cite{robinson1963linear,popov1972relationships,byler1987relation,mayo1990dreiding}, we infer that the strength of single carbon-carbon and carbon-nitrogen bonds should not differ by more than 20\%. So, in the context of the present study, it means that NV centers can be treated similarly to vacancies when one considers their effects on nanoparticles' Raman spectra. However, this issue deserves more in-depth study, for instance with the use of the \textit{ab initio} methods which can shed more light on the peculiarities of carbon-nitrogen bonds of NV centers.

Finally, we note that the case of bond disorder without mass disorder does not require additional consideration. According to Eq.~\eqref{DisParam}, one should use 
\be
  u=-\frac{\delta K}{4m\omega_0} 
\ee
as the disorder strength parameter instead of its counterpart for the mass disorder. Thus, all the aforementioned regimes could also be found for the bond disorder case since there is a correspondence of heavy impurities with weak bonds and vice versa.

\section{Summary}
\label{SDisc}

Here we summarize our findings and discuss their practical applicability. We argue that despite the complex nature of the problem, there is a simple recipe for how to incorporate the disorder-induced peak shift and broadening into a particular Raman spectrum treatment. To do that, one should perform in the Lorentzian-shaped denominator in Eq. \eqref{eq_ramanvianormal} the replacement $\omega_\nu \rightarrow \omega_\nu +\delta\omega(c_{\rm imp})$. In the same manner $\Gamma_\nu = \Gamma_\nu^0 + \Gamma(c_{\rm imp},L)$, where $\Gamma_\nu^0$ absorbs all other mechanisms of phonon damping including spectrometer resolution. For the Phonon Confinement Model, similar substitutions are also in order. It is also pertinent to note that for more precise shift calculations the bulk value for ideal diamond crystal $\omega_0$ is required.

\subsection{Light atoms, vacancies and NV centers}
For light replacement defects (light dopant atoms) and vacancies (including the case of NV center-type impurities) with a concentration $c_\textrm{imp}$ in the range 1-10\%, additional redshift takes place
\be \label{VacRel1}
  \delta \omega(c_\textrm{imp}) \approx -2.6~\mathrm{cm}^{-1} \cdot \frac{c_\textrm{imp}}{1\%}.
\ee
And for vacancies-induced broadening
\be \label{VacRel2}
  \Gamma(c_\textrm{imp},L) \approx 1.5~\mathrm{cm}^{-1} \cdot \frac{c_\textrm{imp}}{1\%} \cdot \frac{3~\textrm{nm}}{L}.
\ee

For lower concentrations, both shift and broadening are comparable with an instrument error and do not contribute to Raman spectra according to accumulated massive experimental data on nanodiamonds. For higher $c_\textrm{imp}$, the developed theory is inapplicable.

\subsection{Heavy impurities}

In the limiting case of heavy impurities (at least 1.5 heavier than the matrix atom) the disorder-induced peak redshift and broadening saturate as functions of $\delta m$
\be \label{dw2333}
  \delta \omega(c_\textrm{heavy}) = - 1.0~{\rm cm}^{-1} \cdot \frac{c_\textrm{heavy}}{1\%} 
\ee
and
\be \label{gamma2333}
   \Gamma(c_\textrm{heavy},L) \approx 0.5~\mathrm{cm}^{-1} \cdot \frac{c_\textrm{heavy}}{1\%}  \cdot \frac{3~\textrm{nm}}{L}
\ee
For impurities with masses closer to the matrix atom mass, the effect is significantly less pronounced being proportional to $\delta m/m$. In this context, another important conclusion is that isotopic disorder in diamonds cannot yield a broadening of the order of 1~cm$^{-1}$ in diamond nanometer-sized particles in all ranges of concentrations.


\subsection{Resonant cases}

As we show above, in resonant cases the effect of disorder on position and broadening of nanocrystalline nanoparticle Raman peak is drastically enhanced. For additional disorder-induced peak shift $\delta \omega$ in nearly resonant conditions (impurity mass approx. 78\% of matrix atom mass or defect bond rigidity approx. 1.4 with respect to ``pure'' bond force constant) one can use the estimate $\delta \omega \propto \Gamma$, see Eqs. \eqref{dw2} and \eqref{gamma2} with the coefficient or the order of unity (disorder-induced peak shift is now positive). Importantly, both disorder-induced redshift and broadening are insensitive to nanoparticle size in that case and can serve as fitting parameters for experimental data.

\section{Conclusion}
\label{SConc}

To conclude, in the present research, we study the intimate relationship between the optical phonon line broadening due to point-like lattice disorder and the Raman peak shift with respect to the pure material in crystalline nanoparticles. Considering widely used diamond particles as a representative example, we show that various regimes of phonon line broadening and Raman peak shift exist. In simple words, they correspond to light, heavy, and resonant impurities. Importantly, in all those regimes for nanometer-sized particles (the heavy impurities concentration should not be very small), the broadening and the shift are of the same order of magnitude. We derive the formulae for the corresponding ratios which, as a rule, are dependent on the particle size, and discuss them using the results obtained in the framework of the self-consistent T-matrix approach. Special attention is paid to the prominent in diamonds NV centers. They are shown to lead to a similar with vacancies effect on the optical phonon modes. 

From the applied point of view, the obtained results are important to properly account for both disorder-induced effects when treating experimentally measured Raman spectra. Usually, the phonon line broadening parameter is considered an adjustable one. We show that assuming some dominant sort of impurities (e.g., NV centers in nanodiamonds) we can relate the broadening and the peak shift thus making a description of the corresponding experimental spectra more accurate.

\begin{acknowledgments} 

S.V.K. acknowledges IBS Young Scientist Fellowship (IBS-R024-Y3). O. I. U. acknowledges financial support from the Institute for Basic Science (IBS) in the Republic of Korea through Project No. IBS-R024-D1.

\end{acknowledgments}

\bibliography{bib}

\end{document}